\documentclass[a4]{article}
\usepackage{amssymb}
\begin{document}
\title{Distance to the drip lines}
\author{Alejandro Rivero \thanks{EUPT (Universidad de Zaragoza), 44003 Teruel, Spain
{\tt arivero@unizar.es}}}
%\pacs{21.30.-x, 14.80.Bn}??? viejos
%\keywords{neutron drip line, proton dripline}
\maketitle

\begin{abstract}
It can be found that with the adequate measure, 
the beta stability line is equidistant from neutron and proton drip lines. We explore this
fact and its predictive potentiality in the simplest case, the classic liquid drop formula.
\end{abstract}

For a nucleus of atomic number $A=N+Z$ in the beta stability line, we can consider the corresponding 
nuclei $(Z-k,N)$ in the proton drip line and $(Z,N-k')$ in the neutron drip line, with respective
masses $A_p, A_n$. The main mass models in the market (eg, from \cite{uno}) predict a very
small difference $A_p-A_n$, which even becomes zero in isolated points under the action of
microscopical corrections.  

We have studied this difference for the classical Weizs\"acker formula\cite{bethe}, 
$$E_b=a_1 A - a_2 A^{2/3} - a_3 {Z^2 \over A^{1/3}} - a_4 {(A-2Z)^2 \over A}$$

An analytical -even if very large- expression can be given if instead of taking $A$ as the
independent variable, we fix the mass $A_0$ in the drip lines. Then solving the second degree
equation in the proton drip line
$$M[Z, A_0] - M[Z - 1, A_0 - 1] - m_z=0$$
and the third degree one in the stability line (we take $m_p-m_n\sim 0$ but it is not necessary)
$$
Z={2 a_4 A \over a_3 A^{2/3} + 4 a_4}
$$
we can get the corresponding mass A and proton and neutron numbers $Z, N (=A-Z)$ of the
stable nucleus. We compare this neutron number with the one got from the
neutron drip line equation
$$M[A_0 - N_0, A_0] - M[A - N_0, A_0 - 1] - m_n =0.$$
The difference $d(A_0)=N-N_0$ results a very convenient function to input in
a numerical-analytical program such as, for instance, {\it Mathematica}, because we can plot 
dependences with any of the four free parameters of the model, as well as mixed plots 
$d(A_0,a_i)$ or $d(a_i,a_j)$. It is specially relevant to check the dependence in $a_3$,
because it has a natural minimum for the zero value, but it is not uniformly increasing;
there is a second minimum in the $\sim 1$ MeV  area, but this one has also a
dependence on $A_0$ so we can no expect it to coincide exactly with the usual
value $a_3 = 0.711$ MeV. Still, this minimum can be interpreted as the cause of our 
equidistancy.

For the usual values
$$a_1 = 15.75 \mbox{MeV}, a_2 = 17.8 \mbox{MeV}, a_3 = 0.711 \mbox{MeV}, a_4 = 23.0 \mbox{MeV},$$
it can be seen
that $d(A_0)$ gets the higher value for $A_0\sim 300$; it is only -0.815 
when proton and neutron masses are equal, and this maximum discrepancy moves by only
about two units when proton and neutron masses are given different value, so for
simplicity one can find convenient to keep with $m_p-m_n\sim 0$ as we do here. 

The function is not linear, so for mid-range masses the difference is appreciably smaller. Generically we
can say that the equidistance property $k=k'$ with different proton and neutrons masses holds within a 2\%.

As a numerical experiment, and to test how strong a rule is the equidistance to the driplines, we
measured the square discrepancy at four different values, using an averaging function
$$\mbox{mean}= d(50)^2+d(100)^2+d(150)^2+d(200)^2,$$
and we took  the bold bet of searching for
 minima\footnote{Mathematica 5.x users beware: the default method in {\tt FindMinimum} notices the
sum of squares and it fails; you must use any of the older methods.} in the allowed parameter
space of $a_1... a_4$.

After discarding unphysical or non-convergent zones, we explored a detailed area 
for $a_1=6..18, a_2=0..40, a_3=0.3..1.2, a_4=5..45$. We seed the algorithm with values in this
area and then we ask it to descend numerically into a local minimum. Note that smaller
values of $a_3$ cause the algorithm to descent into the trivial $a_3=0$ case.
The results are seem in the table. The stability line depends only of $a_3/a_4$, so it is a good
indication to get both increasing, even if below the empirical value. 
It can be said that we are always out, in the four parameters, by a sensible percentage,
but if one takes into account that no empirical input is used in the calculation\footnote{except, if you
want, the value of the atomic mass unit}, the exploration is successful enough. Besides, we have chosen
the averaging function in a very arbitrary way. This choosing is 
 surely more important than the seed. For the usual values, $d(A_0)$ has three zeros besides the trivial, about
$A=25, 600, 2000$. This is very generic, and it means that up to three discrete values can be adjusted in an
unnatural manner. The averaging formula must evaluate a number of points enough to be sure that no zeros happen
in the range of validity of the mass formula, say $A=30...300$. From the shape of $d(A_0)$ in this range it
seems that a weighting proportional to the binding energy of $A_0$ (say, in 
the stability line) could be adequate, but it is
still an ad-hoc suggestion, as any function with the same shape would do the
same role.
 
 Experiments can
be done with variations in only three, two or one parameters, but then the extant ones must be
fixed with other methods, beyond the purpose of this brief note. Contrary to the others, the
 parameter $a_4$ seems not be very able to vary, as it stops about two or three units from the 
 seed. But fixing it to 23.0 does not improve the match. 

As we have said, the equidistance property can be noticed in most models of nuclear masses, and
our function $d(A_0)$, or alternatively any measure of the discrepancy between $k$ and $k'$, is an
 interesting parameter to consider when studying the properties of a mass model.

An explanation of this property should be that really some important mass
dependent effect is enhanced at the drip lines, so this effect forces mass formulae to
adjust themselves to fit. The effect does not need to happen uniformly in all the 
mass range, it is enough to force the coincidence in isolated points, perhaps the ones
having strong microscopical corrections, as noted above.

Perhaps it is related to the extra nuclear stability coming  from magicity, because
an additional phenomena appears when we consider magic numbers at the drip lines:
we can use nuclear mass to pair Z and N numbers, in such way that a nucleus in
the proton drip line given a magic (or semimagic) Z number correspond to another
nucleus of the same mass in the proton neutron drip line, having again a magic
(or semimagic) N number:

 \begin{tabular}{l || c | c| c  |c |c |c}
 At neutron dripline, N= & 28-30 & 50 & (64) & 82 & 126 & 184 \\
 \hline
 At proton dripline, Z= & 28 & (40)  & 50 & (58) & 82 & 114
 \end{tabular}

 \begin{table}[h]
 \begin{tabular}{ | c || c | c| c |c|}
 mean sqr discrepancy& $a_1$&   $a_2$  & $a_3$ & $a_4$   \\
 \hline
0.00329185&11.1382&18.2911&0.521987&18.9961\\
0.00450291&11.0895&19.869&0.556106&17.1047\\
0.0173419&13.9102&20.5569&1.62424&13.6503\\
0.0158552&14.8943&19.1126&2.07992&14.2993\\
0.0144648&15.4414&18.2103&2.34985&14.7491\\
0.00267294&12.2724&18.3904&0.568557&22.0129\\
0.0021288&13.5535&18.1768&0.606681&26.1328\\
0.00175986&15.0299&17.7684&0.668602&30.3896\\
0.00149186&16.6956&17.0859&0.748912&34.9357\\
0.00108511&18.1636&12.6318&0.812369&41.3469\\
0.00110731&18.2587&12.8536&0.823127&41.2093\\
0.00113735&18.1823&13.6355&0.811909&40.8982\\
0.00212587...(*)&14.2744&16.8097&0.70215&26.5332\\
0.00214133...(*)&13.7113&18.2199&0.628928&25.981\\
 \end{tabular}

 \caption{ \label{math} Some typical predictions from Mathematica' {\tt FindMinimum} explorations
  of  the allowed parameter
 range for  the liquid drop model. To be compared with the usual $a_1=15.75, a_2=17.8, a_3= 0.711,
  a_4= 23.0$. Lines marked with (*) took 15.75 and 17.8 as seed}
  
 \end{table} 
  
By the way: the author was too worried about other phenomena and he
thought the fact here reported were well known. So I must thank M. Asorey by asking why the drip lines I kept
drawing were always equidistant of the stability, and A. Zuker by suggesting to look first in the simplest
model,  Weizsaecker's liquid drop.


\begin{thebibliography}{30}
 

 \bibitem{bethe}
 H. A. Bethe and R. F. Bacher 
{\it Nuclear Physics A. Stationary States of Nuclei}
 Rev. Mod. Phys. 8, 82-229 (1936)
 
{\tt  http://link.aps.org/abstract/RMP/v8/p82}
  
 \bibitem{uno}
 M. Uno,
 {\it Current status of nuclear mass formulae and their predictability}
 RIKEN review, 26 (2000),  38-44
 
 {\tt http://www.riken.go.jp/lab-www/library/publication/review/conts/conts26.html}
 
 \end{thebibliography}
 \end{document}